# Bidirectional propagation of cold atoms in a "stadium" shaped magnetic guide

Saijun Wu, Wilbert Rooijakkers*, Pierre Striehl, Mara Prentiss

Department of Physics, Harvard University, Cambridge, MA, 02138
Division of Engineering and Applied Science, Harvard University, Cambridge, MA, 02138

**Abstract:**

We demonstrate the bi-directional propagation of more than $10^7$ atoms ($^{87}$Rb) around a "stadium" shaped magnetic ring that encloses an area of 10.9 cm$^2$, with a flux density exceeding $10^{11}$ atoms sec$^{-1}$ cm$^{-2}$. Atoms are loaded into the guide from a 2D$^+$ MOT at one side of the "stadium". An optical standing wave pulse is applied to increase the propagation velocity of atoms along the waveguide. The atom sample fills the entire ring in 200 ms when counterpropagating atom sections of the original atom cloud recombine at their initial positions after a full revolution. We discuss a possible extension of this result to a guided atom ring interferometer.

*) Present address: University of Melbourne, School of Physics, Victoria 3010, AUSTRALIA

**Introduction**

There is great interest in exploiting atom waveguides to improve precision measurements. In particular, fully reciprocal guided wave interferometers with large enclosed areas may be able to significantly increase both the accuracy and the sensitivity of atom based rotation sensors. Rotation sensing requires that two halves of the initial wavepacket propagate around a path that encloses an area. If the interferometer is rotating, then the Sagnac effect produces opposite phase shifts in the two wavepackets, where the phase shifts are proportional to both the rotation rate and the area enclosed by the paths traversed by the wavepackets. Measurements of the phase difference between the two wavepackets can then be used to determine rotation, where the sensitivity is proportional to the enclosed area.

In demonstrated atom interferometers, the area $A$ enclosed by the two arms is typically ~10 mm$^2$ ($A = 22$ mm$^2$ in [1]), limited by the splitting angle of the atomic beam splitter. In contrast, in a guided wave interferometer the wave paths may be separated much more. Guided wave interferometers that permit the separated wavepackets to traverse the reciprocal path should provide good common rejection of phase shifts associated with time independent energy variations along the path of the interferometer. Time dependent perturbations will also be cancelled if the time scale for the changes in the perturbation is longer than the time required for the wavepacket to traverse the interferometer, or when they occur in a region of negligible atom density.

It has already been shown that guided atoms can circulate around a closed path with an effective area as large as 4400 mm$^2$ [2]. However, in that experiment the atom sample could not easily be split into two counterpropagating wave packets whose phase difference could be used to determine rotation. Indeed, a major experimental challenge in proposals for guided matter wave interferometers is the realization of coherent beamsplitters [3-6]. As was suggested in [7, 8], optical standing waves may be used instead of the more cumbersome method of using a bifurcation in the guiding potential. Matter wave interferometers using diffraction from optical standing waves as beamsplitter have proved to be efficient and reliable in free space [9-11]. Such beam splitters may also coherently split the atom wavepackets confined by waveguides. If the momentum impacted by the standing wave is along the propagating direction of the waveguide, the resulting atom sample can be used in guided atom interferometers.

It has been shown theoretically that single mode guided atoms can be split with an optical standing wave and interfere on a ring [7]. In that article, the authors carefully examined the influence of the atom-atom collisions, and pointed out that a reliable interference signal can be prepared from either fermions or bosons. For bosons the phase shift is complicated by the mean field interaction in the limit of large density. However, the authors of [7] show that in the Tonks gas regime the phase shift still faithfully represents rotation as a consequence of boson-fermion mapping theory. Similar interference is expected for a dilute atom sample that populates multi modes of the ring waveguide [12]. These proposals require an atom waveguide loop that can be combined with standing wave beamsplitters.

In this paper, we demonstrate a novel type of magnetic waveguide ring that is compatible with the standing wave beamsplitter. The magnetic waveguide ring is composed of two half circles which are connected with two straight parts, thus composing a "stadium" shape. (Fig.1). The atoms are loaded into the waveguide from a 2D$^+$ MOT operated at one straight part of the "stadium", with the same technique described in [13]. After the transfer, an optical standing wave is applied to increase the velocity of atoms along the direction of waveguide. The atoms propagate in opposite directions around the ring and recombine at the original location. We expect a similar device can be used for guided atom interferometry.

**Experiments with a magnetic "stadium"**

Our magnetic waveguide ring is composed of 2D magnetic trap in a stadium shaped loop. The 2D magnetic trap is not generated by current carrying wires alone to avoid a large potential discontinuity at the point where the current enters and leaves the ring. Instead we use the 4-foil magnetic structure [13] to generate the 2D magnetic trap (Fig.1.b), and construct the ring simply by connecting the 4-foil magnetic structure into a loop. Specifically, four stadium shaped rings with sorted size, each individually wound with 42 turns of 400 $\mu m$-diameter Kapton ® wire are assembled to form a closed loop (Fig.1). The foils are 0.5 mm thick and spaced from each other by 3 mm. The foils are made of magnetizable material to create enough magnetic gradient for guiding at a distance ~ 1 mm from the surface [13,14]. Mu metal (80%Ni, 15.5% Fe, 4.5% Mo) is chosen since it can easily be poled by current carrying wires, and has negligable remnant magnetism when the current in the wires is switched off. A linear waveguide based on this type of 4-

foil magnetic structure has been discussed in [13], where we demonstrated that 4-foil ferromagnetic structures can create 2D magnetic traps with field gradient tunable from zero up to 1000 G/cm. Folding a linear ferromagnetic waveguide to the shape of a ring opens a possibility to study the properties of atoms stored in a ring with very tight confinement.

Figure 2 shows the calculated magnetic field strength |B| for the structure. The boundary conditions for the simulation were measured directly with a Gauss-meter while the currents were set to be the same as in atomic experiment, and the calculation is confirmed by comparing the location of magnetic minimum in measurement and simulation. When the current is 800 mA through the smallest two coils and 1.68 A through the largest two coils the B fields from the four mu-metal foils cancel approximately 2.5 mm above the foils (2.2 mm above the mirror surface) near the top of $2^{nd}$ ($2^{nd}$ largest) foil, with d|B|/dr = 100 G/cm at the straight parts. Alternatively, one may apply an external B field to cancel the B field generated by the smallest two foils, thus forming a 2D magnetic trap in only one straight section of the stadium. Experimentally we find the second configuration is more favorable for the operation of a $2D^+$ MOT. In situ loading of cold atoms from the MOT to the first configuration is realized by slowly ramping down the external B field while increasing the currents for the four foils in proper ratio.

It's important to point out that, the resulting "stadium" shaped magnetic field minimum is not a field zero loop. In fact the calculation in Fig.2 indicates only in four points (the center of the straight sections and in the center of the curved sections) does the magnetic field cancel exactly, everywhere else we find a small magnetic field along the loop. Also, the waveguide becomes "looser" for atoms entering the curved section (85 G/cm at the center of the curved parts compared to 100 G/cm at the center of the straight parts). In addition, the minimum is slightly elevated in the curved parts (Fig.2), and hence the atoms have to climb a gravitational potential while entering the curved parts. In the experiment we observe that atoms propagating towards the curved parts are partially scattered back from *both* sides (Fig. 3). This variation in potential would be absent in a waveguide with circular symmetry.

In the first stage of the experiment, a 2D$^+$ MOT is created at one side of the stadium, where the 2D magnetic trap is generated by external magnetic field and the smallest two foils. The cooling light has detuning $\delta$ = -14 MHz to the red of the $F = 2 \rightarrow F' = 3$ hyperfine transition in $^{87}$Rb and repumping was provided by an additional laser locked on the F=1 -> F'=2 hyperfine transition of the D2 line. The $1/e^2$ beam diameters of transverse and longitudinal laser beam are 2 cm and 1.5 cm, respectively. The laser intensities are 8 mW/cm$^2$ for each longitudinal beam and 5 mW/cm$^2$ for each transverse beam. During a period of 5 seconds $2 \times 10^7$ atoms are accumulated from the background vapor ($3 \times 10^{-9}$ mbar), producing a cigar shaped sample of cold atoms that is 3 mm long and 0.5 mm wide. After that the currents for the four foils are ramped up, and the external B field is ramped down to zero in such a way that the magnetic minimum remains at about the same position, but the gradient increases from 5 G/cm to 100 G/cm. The compression is completed in 60 ms. During the first 30 ms the cooling light is kept on, but shifted from $\delta$ = -2.4 $\Gamma$ to $\delta$ = - 4.5 $\Gamma$, and is then extinguished. A circularly polarized standing wave is pulsed for various durations after the cooling light is switched off. The evolution of the atom density along the guide is examined using resonant absorption imaging.

Fig. 3a) shows a sequence of the absorption images at one side of the stadium, after a $\delta = +7\Gamma$ (blue detuned) standing wave with saturation parameter s~5 is pulsed for 300 $m$s (defined as time zero in the following discussion). Combination of a 0.1 Gauss plug field and the circular polarized standing wave optically pumps the atoms to the weak field seeking states. From the absorption image we find that more than $10^7$ atoms are transferred from $2D^+$MOT to the straight part of the magnetic waveguide, an efficiency of more than 50%. The transverse temperature of the magnetically trapped atoms is estimated from the extension of atom sample in waveguide to be about half the Doppler temperature. The mean longitudinal expansion speed of the sample is approximately 20 cm/s. The atom sample expands symmetrically, moves beyond the imaging aperture in 30 ms. As shown in Fig. 4, a "stadium" shaped atom distribution emerges after 200ms of propagation inside the waveguide. Since the CCD chip is 10 mm by 7.1 mm, we realign the absorption image system to observe atoms in different sections of the 11.2 cm long loop while fixing the time delay at 200 ms. The images taken at different positions are then used to re-construct the cloud over the full extension of the magnetic guide.

Fig. 4 provides a static picture of atom distribution around the ring. To confirm that the atoms actually propagate around, we use an on-resonant light to "cut away" part of the atom sample, and observe the propagation of the "cut" via absorption images (Fig. 3b).

At t = 100 ms when the atom sample has filled over 3 quarters of the stadium, an on-resonant standing wave ($1/e^2$ beam diameter: 0.8 cm) is pulsed for 5 ms so that the entire half of the stadium on the side where the atoms were originally loaded is depleted. Absorption images were then taken to study the propagation of the "cut" in time and position. The left column of Fig. 3b shows the edge of the atom cloud propagating clockwise towards the straight section of the waveguide, with speed approximately 25 cm/s. At t' = t – 105 ms = 20 ms after the cutting operation a dilute trace blurs out the previously well defined edges. At this time the fastest atoms from the other curved section of the guide have arrived and start populating the empty region. As time advances the other edge should start to appear moving counterclockwise. This is observed at t'=160 ms (right column of Fig 3b). Again, the edge is not sharp because by this time the fastest atoms that had originally started from the right have fulfilled an entire revolution and penetrated through the atoms from the left to blur out the edges. Nevertheless, the sequence of images shows clearly how the edge from the left moves upwards through the curved section of the guide indicating that the atoms really perform an entire revolution. Notice that the atoms keep revolving in the closed waveguide, only limited by the lifetime of trapped atoms in vacuum, which presently is around 1 second.

**Extension to a guided atom ring interferometer**

Although the current experiment has demonstrated the bidirectional propagation of guided atoms around a loop, various experimental limitations have so far prevented us from exploring the coherence properties in the transport of atoms. In this section we will discuss a possible extension of our experiment towards a guided atom ring interferometer.

The proposed guided matter wave ring interferometer works in pulsed mode, similarly to experiment described above. The atoms are initially localized at one side of the circularly symmetric waveguide ring, tightly confined, and longitudinally subject to optical standing waves which coherently split the guided matter wave into velocity class v+nv$_r$ and v–nv$_r$ [15], where n is an integer, $v_r = \frac{\hbar k}{m}$ is the recoil velocity which is 6 mm/s in the case of rubidium. The atoms are then allowed to propagate along the waveguide. The loop structure of the ring imposes a periodic boundary condition and the atom amplitudes of the two velocity classes recombine at the time when atoms with velocity nv$_r$ traverse the entire ring. Similar to the Talbot-Lau type interferometer [11, 16], the resulting interference patterns from atoms with different initial velocity v add up constructively. Depending on the initial velocity spread of the atom sample, the fringe pattern may extend over the entire ring. The fringe pattern near the original location of atom sample takes advantage of reciprocal symmetric evolution and is robust against static potential variations along the guide. The position information of the resulting interference fringe is related to non-reciprocal perturbations such as rotation, and can be read out in various ways with light fields [11, 12], such as phase recovery using heterodyne detection of back scattered probe light.

The proposed ring interferometer can be used as a gyroscope operating in pulsed mode. A single round operation is composed of three steps: loading and splitting of atoms, propagation around the ring, and detection. The period of operation is limited by $T = \frac{S}{nv_r}$, where S is the circumference of the ring. The sensitivity of the gyroscope to rotation is given by $d\Omega \sim df \dfrac{\hbar}{4mA}$, where $df$ is the resolution of the optical phase of backscattered probe light out of $2p$, $A$ is the area covered by the ring. Different from a gyroscope working in continuous mode, the bandwidth of a gyroscope working in pulsed mode is limited by the operation period T, which requires that the rotation rate of the system does not exceed the "free rotation rate" $\Omega_f \sim 2p \dfrac{\hbar}{4mA}$ for reliable readout. This can be achieved by mounting the whole system on a platform that is actively stabilized via an optical gyroscope, which keeps the angular velocity of the system smaller than $\Omega_f$ and gives a rotation readout that is periodically corrected by the atomic gyroscope operation.

The combined gyroscope system will possess the large bandwidth of an optical gyro and the high sensitivity of an atomic gyroscope. Similar to the analysis in [17, 18], in the shot noise limit, the optical phase of a back scattered probe light from a matter wave grating composed of N non-degenerate atoms can be resolved to the order of $df \sim \dfrac{1}{\sqrt{N}}$. As an example, let us consider a circular waveguide ring with radius R=1 cm. Take $df=10^{-3}$ in shot noise limit if $N=10^6$ atoms contribute to coherent backscattering. Consider the splitting order n=11, we have: T= 0.5 second, $d\Omega = 8 \times 10^{-6} \Omega_E$, $\Omega_f = 0.05 \Omega_E$, where $\Omega_E$ is the rotation rate of the earth. Compared with a free space version, the guided atom gyro system proposed here has a larger sensing area while still being relatively compact. The sensitivity and accuracy of the system can be boosted further by combining several atomic gyros on the same rigid platform sharing the same optical setup, where the rings may be loaded sequentially permitted nearly constant rotation measurements even if each individual ring is operated in a pulsed mode.

In the proposed guided atom ring interferometer, small variation of guiding potential along the waveguide is tolerable due to the reciprocal symmetry. Potential variation comparable or even larger than the kinetic energy of the guided atoms would certainly degrade the interference. For the "stadium" shaped magnetic assembly demonstrated here, the variation of the guiding potential can in principle be reduced to arbitrarily small level by reducing the separation of the 4-foils, and carefully adjusting the separation of 4-foils along the curved parts. The variation of guiding potential along a circular magnetic waveguide [19, 20] should be much smaller, a result that has been confirmed by preliminary measurements on such structures.

We emphasize that the single mode guiding is not necessary for the atomic interferometer proposed here. A tight waveguide that separates the guided motion of atoms from the transverse motion is compatible enough with the principle of the interferometer discussed above. Actually, compared to a guided atom interferometer configuration with non-reciprocal paths [3-6], we expect the interferometer proposed here to be more tolerant to coupling of the guided motion with the transverse motion. The tolerance to multi-mode operation allows the proposed atom interferometer to work with dilute instead of quantum degenerate atom gases, thus reducing the complications from collisional phase shifts [7].

To conclude, we have demonstrated bidirectional propagation of $10^7$ cold atoms in a novel magnetic waveguide ring. We propose a guided atom ring interferometer, which combined with an optical gyro may make a precise and compact rotation sensing system.

Acknowledgement

The authors are indebted for valuable suggestions by Mukund Vengalattore on the optimization of trap loading, and valuable discussions with Yanhong Xiao, Scott Sanders, and Richard Conroy. WR acknowledges support from the Commonwealth of Australia DEST Contract CG02-0135. This work is supported by U.S. Department of the Army, Agreement Number DAAD19-03-1-0106.

[19] A curved 2D$^+$ MOT were successfully demonstrated at the curved part of the "stadium" magnetic assembly here. We expect the loading of atoms to a circular waveguide would have similar behavior.

[20] In the tight waveguide limit, momentum exchanged between light field and the guided atom is the projection of the photon momentum along the guiding direction, thus in a circular waveguide the misalignment between the light field and a curved waveguide end up in a broadening of the momentum exchange.

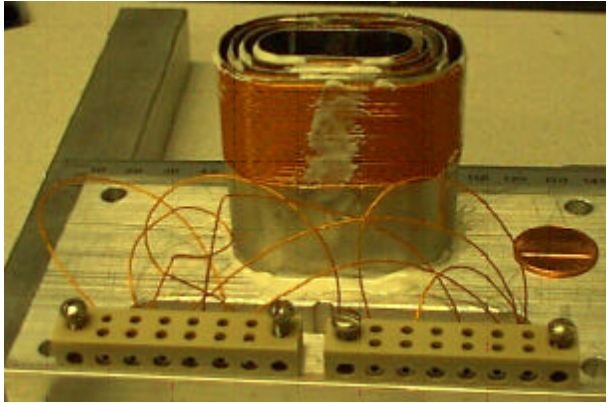
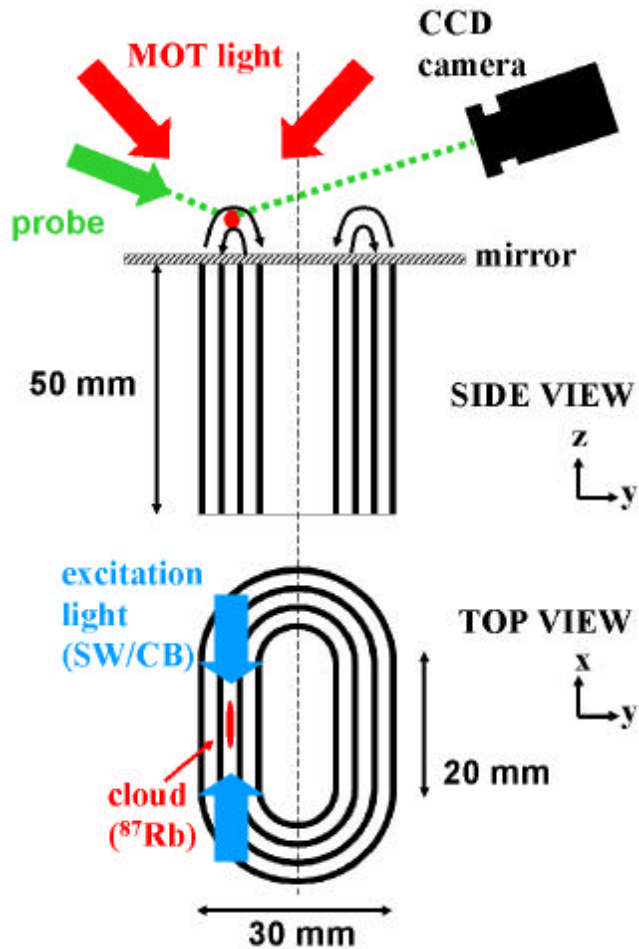

Figure 1 (Color Online): Top: Photograph of the stadium-assembly (without mirror). Kapton ® wires are wrapped around each foil and attached to the mu-metal with Torr Seal glue ®. Bottom: top view and side view of the experimental setup. Same laser beam is used for standing wave (SW) operation ($\delta = +7\Gamma$) and as a cutting beam (CB, on-resonance).

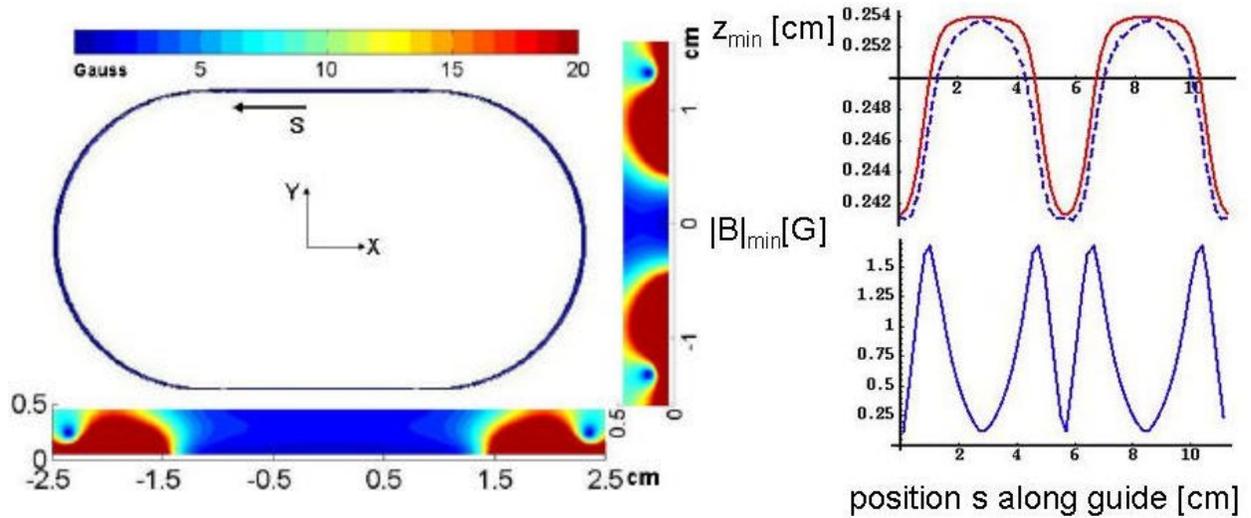

Figure 2 (Color Online): Left: Top view of the isopotential surface of calculated magnetic field strength |B|=2 Gauss for the stadium assembly. Field strength is also shown in y=0 and x=0 planes, coinciding with the long and the short axis of the stadium respectively. Right: Results of a numerical minimum search routine along the circumference s of the guide (see text). Right Up: distance from mu-metal surface, solid line for the magnetic minimum, dashed line for the potential minimum including gravity, assuming $M_F$=2 atoms. Right Bottom: field strength in the magnetic minimum along the guide.

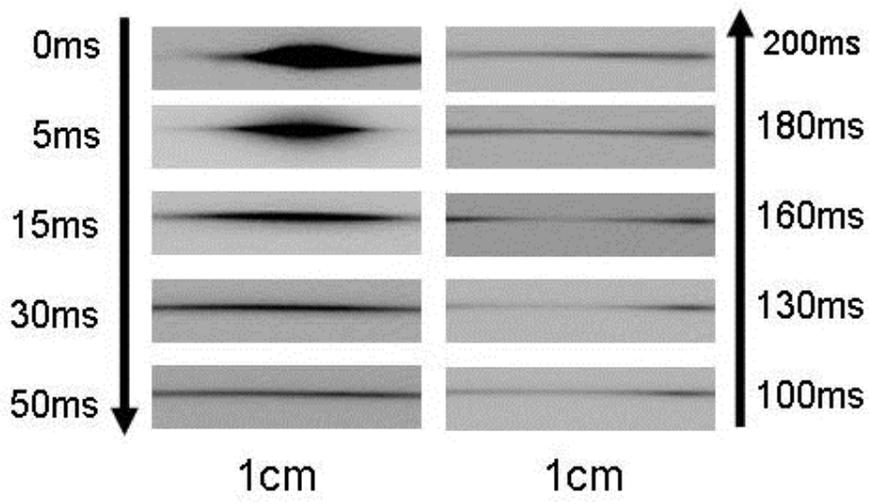

Figure 3: a) Evolution of the atom cloud in the straight part of the stadium after loading at t = 0 ms. Two groups of atoms which make 30% of total atom number come back at around t=160 ms, due to back-scattering from the curved parts.

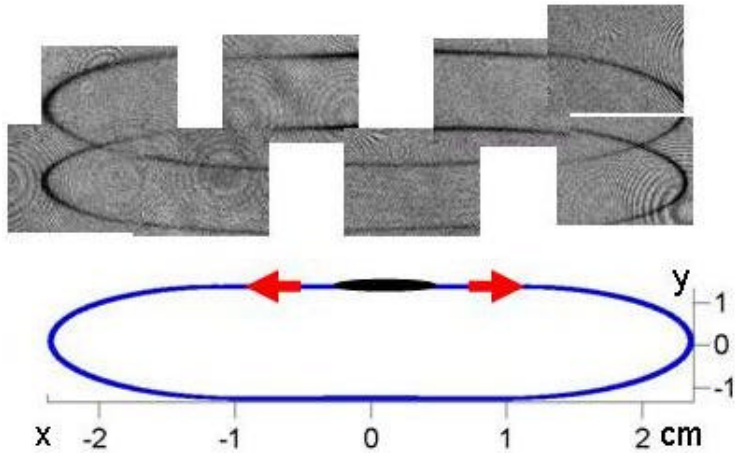

Figure 4: Top: Sequence of absorption images of trapped atoms along the circumference of the stadium after 200 ms of free evolution. Observation angle (~18°), magnification (1:1) and grayscale of each image may vary slightly. Mirror image is also visible. Bottom: Calculated isopotential surface of |B|=2 Gauss, the observation angle is set to be 18°. The initial distribution and propagation directions of atom sample are also indicated.

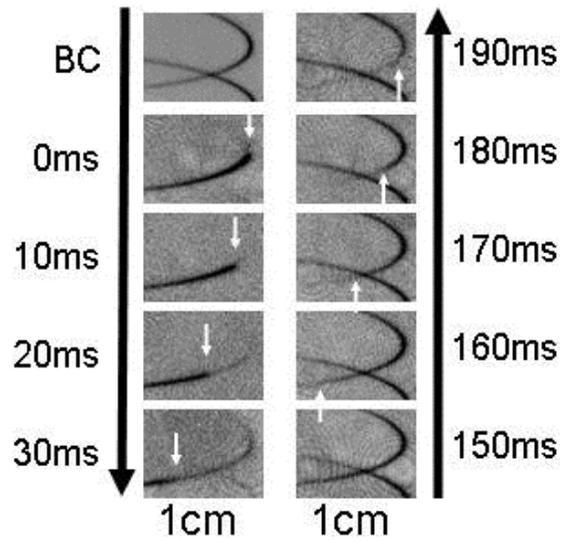

Figure 5: Absorption images near the right curve of the stadium. "BC" = "Before Cut", taken 100 ms after SW operation. The images taken at t' = t – 105 ms (0 ms to 30 ms after cut) show the edge of the cloud propagating "down-left" (clockwise). The images from t'= 150 ms to 190 ms show the edge propagating "right-up" (counterclockwise). This indicates bidirectional propagation around the entire ring. White arrows were added to clarify the edge of the atom cloud.